\documentclass[]{aa}

\usepackage{natbib}
\usepackage{graphicx}
\usepackage{txfonts}
\usepackage{array}

\bibpunct{(}{)}{;}{a}{}{,} 

\newcommand{\grb}{GRB\,021004}
\newcommand{\etal}{et~al.}

\newcommand{\Q}{\ensuremath{Q}}
\newcommand{\U}{\ensuremath{U}}

\newcommand{\chisq}{\ensuremath{\chi^2}}

\begin{document}

\title{Variable polarization in the optical afterglow of \grb\thanks{Based on observations made with the Nordic Optical Telescope; based on observations collected at the European Southern Observatory, Chile, by GRACE (Gamma-Ray Burst Afterglow Collaboration at ESO), under programme 70.D-0523(A)}}

\author{E.~Rol\inst{1}, R.~A.~M.~J.~Wijers\inst{1}, J.~P.~U.~Fynbo\inst{2,3}, J.~Hjorth\inst{3}, J.~Gorosabel\inst{4,5}, M.~P.~Egholm\inst{6,2}, J.~M.~Castro~Cer\'{o}n\inst{7}, A.~J.~Castro-Tirado\inst{4}, L.~Kaper\inst{1}, N.~Masetti\inst{8}, E.~Palazzi\inst{8}, E.~Pian\inst{8,9}, N.~Tanvir\inst{10}, P.~Vreeswijk\inst{11}, C.~Kouveliotou\inst{12}, P.~M{\o}ller\inst{13}, H.~Pedersen\inst{3}, A.~S.~Fruchter\inst{5}, J.~Rhoads\inst{5}, I.~Burud\inst{5}, I.~Salamanca\inst{1}, E.~P.~J.~Van~den~Heuvel\inst{1}}

\offprints{E. Rol (\textit{evert@science.uva.nl})}

\institute
{
  Astronomical Institute, University of Amsterdam,
  Kruislaan 403, 1098 SJ Amsterdam, The Netherlands
  \and
  Department of Physics and Astronomy, University of Aarhus,
  Ny Munkegade, DK-8000 {\AA}rhus, Denmark
  \and
  Astronomical Observatory, University of Copenhagen,
  Juliane Maries Vej 30, DK-2100 Copenhagen, Denmark
  \and
  Instituto de Astrof\'{\i}sica de Andaluc\'{\i}a,
  c/ Camino Bajo de Hu\'{e}tor 24, E-18.008 Granada, Spain
  \and
  Space Telescope Science Institute,
  3700 San Martin Drive, Baltimore, MD 21218, USA
  \and
  Nordic Optical Telescope,
  Apartado de Correos, 474, E-38700 Santa Cruz de La Palma (Tenerife), Spain
  \and
  Real Instituto y Observatorio de la Armada, Secci\'{o}n de Astronom\'{\i}a,
  11.110 San Fernando-Naval (C\'{a}diz), Spain
  \and
  Istituto di Astrofisica Spaziale e Fisica Cosmica, Sezione di Bologna, CNR,
  Via Gobetti 101, I-40129 Bologna, Italy
  \and
  INAF, Astronomical Observatory of Trieste, via G.B. Tiepolo 11, I-34131
  Trieste, Italy
  \and
  Department of Physical Sciences, University of Hertfordshire,
  College Lane, Hatfield, Herts AL10 9AB, UK
  \and
  European Southern Obseratory,
  Alonso de Cordova 3107, Vitacura,
  Casilla 19001, Santiago 19, Chile
  \and
  NASA MSFC, SD-50
  Huntsville, AL 35812, USA
  \and
  European Southern Observatory,
  Karl-Schwarzschild-Stra{\ss}e 2, D-85748, Garching bei M\"{u}nchen, Germany
}

\date{}

\authorrunning{Rol \etal}
\titlerunning{Variable polarization in \grb}
%

\abstract{

We present polarimetric observations of the afterglow of gamma-ray burst (GRB) 021004, obtained with the Nordic Optical Telescope (NOT) and the Very Large Telescope (VLT) between 8 and 17 hours after the burst. Comparison among the observations shows a 45 degree change in the position angle from 9 hours after the burst to 16 hours after the burst, and comparison with published data from later epochs even shows a 90 degree change between 9 and 89 hours after the burst. The degree of linear polarization shows a marginal change, but is also consistent with being constant in time. In the context of currently available models for changes in the polarization of GRBs, a homogeneous jet with an early break time of $t_b \approx 1$ day provides a good explanation of our data. The break time is a factor 2 to 6 earlier than has been found from the analysis of the optical light curve. The change in the position angle of the polarization rules out a structured jet model for the GRB.

\keywords{gamma rays: bursts -- polarization -- radiation mechanisms: non-thermal}

}

\maketitle

%

\section{Introduction}

The generally accepted source of gamma-ray burst (GRB) afterglow emission is synchrotron radiation, produced when the initial relativistic blast wave hits the circumburst matter and starts radiating \citep{rm:1992,pr:1993,mr:1997,wrm:1997,wg:1999} . Synchrotron radiation is highly polarized, up to 75\% \citep{rl:1979}, and polarization has indeed been measured for 6 GRB afterglows \citep{wijers:1999,covino:1999,rol:2000,bjornsson:2002,covino:gcn1431,covino:gcn1498,bersier:2003a,masetti:2003}. See also the reviews by \citep{bjornsson:2003} and \citep{covino:2003b}. These measurements and obtained upper limits \citep{hjorth:1999, covino:2002} show that the level of polarization is generally small, presumably because the intrinsically high polarization is averaged out to the few percent observed \citep{gw:1999,gruzinov:1999,ml:1999}.

If the outflow of the blast wave is collimated into a jet \citep{rhoads:1997,rhoads:1999,sph:1999}, several models predict changes in the degree of linear polarization from a few up to 30\% \citep{sari:1999,gl:1999,rossi:2002}.
So far, only hints for these variations have been seen \citep{rol:2000}, mainly because of the low polarization values that are measured and the difficulties involved in obtaining a time series of accurate polarization measurements of GRB afterglows. One exception is possibly GRB\,020405, for which \citet{bersier:2003a} find a large variation in the degree of linear polarization within a short time interval \citep[see also][]{covino:2003a}, which cannot be reconciled with any current model.

\grb\ was localized with the wide-field X-ray Monitor (WXM) on board the High-Energy Transient Explorer-II (HETE-II) with an initial positional error of 10$\arcmin$. The position was immediately issued to the community, which allowed the rapid discovery of its afterglow with the Oschin/NEAT robotic telescope \citep{fox:gcn1564}.

The afterglow light curve is well covered and shows some deviations from a standard power-law decay, for which various explanations have been offered, such as variations in the burst energy or variations in the density of the surrounding medium \citep[see for example][]{lazzati:2002,heyl:2003,nakar:2003,dado:2003}. \citet{holland:2003} measure a break in the light curve between 3.5 and 7 days after the burst. The redshift of the afterglow plus host galaxy was determined to be $z=2.33$ \citep[see for example][]{moller:2002}, while the spectrum consists of a complex of absorption systems \citep{salamanca:gcn1611,mirabal:gcn1618,moller:2002} at the redshift of the host.

Polarization measurements were obtained by various groups \citep{covino:gcn1595,wang:2003,rol:gcn1596}.
Here, we report on early polarimetric observations obtained by our group with the Nordic Optical Telescope (NOT) at the Observatorio del Roque de los Muchachos on the Canary Islands, and later with the Very Large Telescope (VLT) at the European Southern Observatory (ESO) in Chile.

%

\section{Data reduction and analysis}

Polarimetric observations of the afterglow of \grb\ at the NOT were performed from October 4.859 UT until 4.908 UT ($\sim$8 to 10 hours after the burst) with the Andalucia Faint Object Spectrograph and Camera (ALFOSC), using two calcite plates at different orientations and a Bessel R filter. The calcite plate yields two overlapping images of the field-of-view (FOV) of the telescope, separated by about 15 arcsec. One image allows for the measurement of the ordinary ray, the other image gives the extra-ordinary ray, which together allow for the measurement of one of the linear Stokes parameters.
For each observation, the orientation of the calcite plate was either 45 degrees or 90 degrees. We obtained 3 pairs of exposures, each exposure with an integration time of 600 seconds, which allows the determination of Stokes \Q\ (0/90 degrees polarization orientation) and \U\ (45/135 degrees) for each pair.

The polarimetry observations with the FOcal Reducer/low dispersion Spectrograph (FORS\,1) at the VLT Antu were performed a few hours later, from October 5.151 UT until 5.196 UT. We used a Wollaston prism with a rotatable half-wave plate at four different angles. The images do not overlap but are separated by a mask covering half the FOV. Each angle allows a measurement of both the ordinary and extra-ordinary ray, which in turn allows the determination of the Stokes \Q\ and \U\ parameters. The broad-band filter applied here was Bessel V. The observations consisted of three sets of four exposures, with an integration time of 120 seconds for each of the first four exposures, and an integration time of 300 seconds for each exposure in the last two sets.

All data were reduced using the IRAF\footnote{IRAF is distributed by the National Optical Astronomy Observatories, which are operated by the Association of Universities for Research in Astronomy, Inc., under cooperative agreement with the National Science Foundation.} software suite. The images were first bias subtracted. The flatfielding of the ALFOSC images was performed with the orientation of the calcite plate for the flatfield identical to that of the target image. 
The FORS\,1 images were flatfielded using flatfields without the Wollaston prism and half-wave plate. Artefacts introduced by the prism or half-wave plate can be corrected for using two observations, one with the half-wave plate oriented at either 0 or 22.5 degrees, and one with the half-wave plate oriented at 45 degrees difference.

The ALFOSC images show filter scratches which could not be completely corrected for by flatfielding. To allow for the detection of errors introduced by such artefacts, the position of the afterglow on the CCD was slightly offset from the centre on the third set of observations, while it was centered for the first two sets. Observations of standard stars verified that a source positioned in the centre gives the correct polarization.

To measure the flux, we used aperture photometry on both the ALFOSC and FORS\,1 images, where the apertures sizes were adapted to the measured seeing. Due to the small FOV of ALFOSC in polarimetric mode, only a few stars could be measured, and it was not possible to derive an accurate point-spread function (PSF) for the two images and perform PSF photometry. 

For the determination of the Stokes parameters, only the relative flux of the ordinary and extra-ordinary ray is required. In the ALFOSC case, however, the calcite plate projects images for the ordinary and extra-ordinary ray differently, which results in the PSFs being different for the two images, and, for small apertures, the incorrect ratio of light to be measured. 
To obtain the most accurate as well as precise result, we chose the smallest aperture size for which the resultant Stokes parameters are still consistent with those measured using larger apertures.
An aperture with a radius equal to $\sim 10$ pixels, or 1.5 times the seeing, provided the best results. This aperture is somewhat larger than the one used by \citet{bjornsson:2002}, who performed polarimetric observations of GRB\,010222 with ALFOSC, but their observations were performed with better seeing ($\sim 0\farcs9$ compared to $\sim 1\farcs2$ for our observations).

Calibration and verification of the procedures was done with both zero polarization standard stars, to check for any instrumental polarization, and high polarization stars. For ALFOSC we used BD\,+28\degr4211 and BD\,+32\degr3739 as zero polarization stars, and HD\,204827 as high polarization standard star. For FORS\,1, BD\,$-12$\degr5133 was used as high polarization standard star. 

\begin{figure}
\centering
\includegraphics[width=\columnwidth,angle=0]{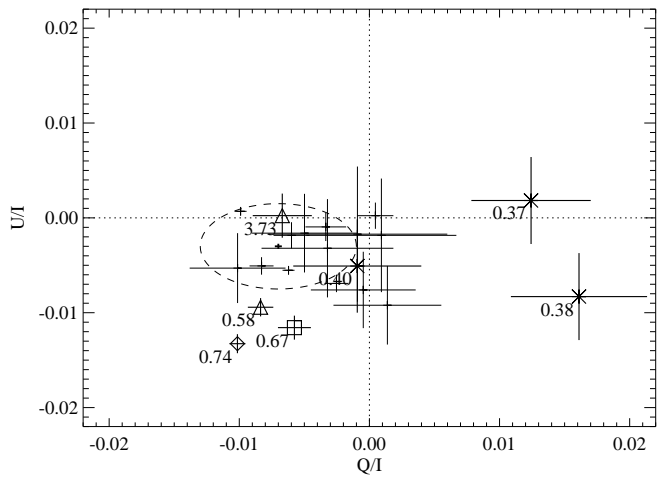}
   \caption{A plot of the Stokes vectors (\Q,~\U) and their error bars (1-$\sigma$ of the afterglow at different epochs, and of the field stars as determined by the FORS\,1 observation. The afterglow is annotated with the time after the burst in days. The ALFOSC measurements are denoted with an asterisk and the FORS\,1 measurement with a square. We have also included data from the \citet{covino:gcn1595, covino:gcn1622} (denoted with a triangle) and from \citet{wang:2003} (diamond). The field stars are plotted without symbols. The dashed ellipse encloses the surface which contains 68\% of the field stars, according to their measured spread. The weighted mean (\Q,~\U) value of the field stars is represented by the thick cross, where the size of the cross indicates the error in the mean. We have used the spread in the field stars as the error in the ISM induced polarization, rather than the error in the weighted mean. }
      \label{figure:quplot}
\end{figure}

We performed aperture photometry on the FORS\,1 data and verified the results by PSF photometry. We used an aperture of 1.5 times the seeing. As there is no difference for FORS\,1 in the PSF between the images of the ordinary and the extra-ordinary ray, this resulted in modest errors. We used the other stars in the FORS\,1 FOV to calculate the polarization introduced by interstellar matter, assuming that the net polarization of the field stars is zero and that the largest fraction of interstellar matter is between these field stars and the observer. Any resultant polarization is then caused by the interstellar matter, and was found to be $P = (0.58 \pm 0.33)\%,\ \theta = (105 \pm 16)\degr$
Since the spread in the (\Q,~\U) values of the field stars is rather large, as can be seen in Figure \ref{figure:quplot}, we used this spread as the error in the above values, instead of the error in the weighted mean of the field stars, which is much smaller.

After calculation of the Stokes parameters, we corrected for polarization induced by the ISM. The degree of linear polarization and its position angle were calculated from both ISM-corrected and uncorrected Stokes parameters, after which the polarization degree was corrected following \citet{wk:1974} and \citet{ss:1985} for bias resulting from the fact that $P$ is a definite positive quantity.

\section{Results}

For both ALFOSC and FORS\,1, we obtained 3 sets of polarization measurements. The results from FORS\,1 are constant from one set to another, showing no variations on this short time scale (15 to 16 hours after the burst). We therefore used the summed image to obtain a higher signal to noise ratio and one final result. For ALFOSC, the three sets do not show evidence for such constancy, and we calculated the results separately (see Table \ref{table:results}). Since the polarization during the ALFOSC observation could have been variable, the assumption of constant polarization, needed to calculate $P$ and $\theta$ from the separate measurements with the calcite plate at 45 and 90 degrees, is not valid. To see whether the results in Table \ref{table:results} are still representative of the polarization at the times of observation, we paired the observations differently, obtaining two more sets. The resulting values for $P$ and $\theta$ are consistent with the first two sets of the original values, that is, they are intermediate values. 
The results for the third set are not consistent with the previous two results and either show a very rapid change in the polarization, or are due to an artefact in the data. The latter could result from to the aforementioned defects in the filter, caused by the afterglow positioned at a faulty position on the CCD (the first two measurements do not suffer from this, as outlined above).

\begin{table}
  \caption[]{Polarimetric results} \label{table:results}
  \begin{tabular}{l@{\,\,}c@{\,\,}c@{\,\,}c@{\,\,}c}
    \hline
              & ALFOSC & ALFOSC & ALFOSC & FORS\,1 \\
              & set 1  & set 2  & set 3  & sum \\
     $\Delta$t (days) & 0.37 & 0.38 & 0.40 & 0.67 \\
    \hline
     \multicolumn{5}{l}{\textit{no ISM correction}} \\
     P (\%)       & $1.17 \pm 0.46$     & $1.73 \pm 0.51$    & $0.15 \pm 0.49$     & $1.29 \pm 0.13$ \\
     $\theta$ (\degr) & $184.2 \pm 11.4$ $^{\mathrm{I}}$ & $166.4 \pm 8.1$ & $129.8 \pm 41.0$ $^{\mathrm{II}}$ & $121.8 \pm 2.8$ \\
     \multicolumn{5}{l}{\textit{with ISM correction}} \\
     P (\%)      & $1.72 \pm 0.56$     & $2.09 \pm 0.60$    & $< 0.59$        & $0.75 \pm 0.42$ \\
     $\theta$ (\degr) & $ 187.7 \pm 8.3$ $^{\mathrm{I}}$  & $173.0 \pm 7.9$ & $-$                     & $132.5 \pm 13.9$ \\
    \hline
  \end{tabular}
  \begin{list}{}{}
    \item[$^{\mathrm{I}}$] We added 180 to the value of the angle for clarity.
    \item[$^{\mathrm{II}}$] The very low value for the degree of linear polarization makes the value for position angle very insecure, which is reflected in the large error. See also the text for comments on this data point.
  \end{list}
\end{table}

When comparing the ALFOSC data with the FORS\,1 data, we assume there is no significant difference in the polarization due to the different filters (V and R) we have used. \citet{wang:2003} mention a wavelength dependent change in the polarization for their spectropolarimetric measurements, but this only occurs below $\approx 405$ nm and would not affect our comparison.

\begin{figure}
\centering
\includegraphics[width=\columnwidth,angle=0]{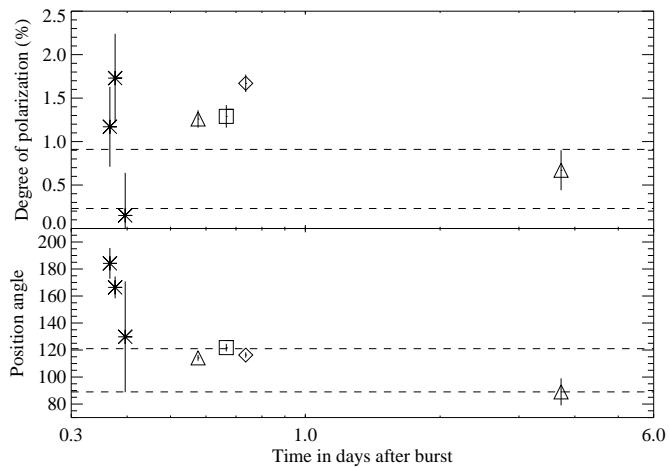}
   \caption{A plot of the degree of linear polarization $P$ and the position angle $\theta$ as function of time. ALFOSC measurements are indicated with an asterisk, the FORS\,1 measurement with a square. We have included the data from \citet[][triangles]{covino:gcn1595,covino:gcn1622} and from \citet[][diamond]{wang:2003}. 
The dashed lines give the $\pm$1-$\sigma$ ranges for the ISM polarization.
}
      \label{figure:pthetaplot}
\end{figure}

We have plotted the resultant degree of linear polarization and position angle as a function of time in Figure \ref{figure:pthetaplot}, where we have also included the results by \citet{covino:gcn1595,covino:gcn1622} and \citet{wang:2003}. From the data uncorrected for ISM polarization, we see rapid changes in the degree of polarization between 8 and 10 hours after the burst. The polarization measured by FORS\,1 could be entirely due to the ISM polarization, since the spread in the ISM polarization is rather large, as also remarked by \citet{covino:gcn1622} and \citet{rol:gcn1596}. However, the change in the polarization from our FORS\,1 data point to the one measured by \citet{covino:gcn1622} shows that the polarization is at least partially intrinsic to the afterglow.

A change in the position angle and degree of linear polarization is entirely due to the afterglow, assuming that the polarization of the field stars and ISM is constant in time. The first two ALFOSC measurements are consistent with having a constant position angle;
the later FORS\,1 point shows a change by about 45\degr\ in the position angle at a 5 sigma level. Inclusion of the measurement by \citet{covino:gcn1622} even indicates a change of about 90\degr\ from 9 to 89 hours after the burst.
For clarity, we have plotted the \Q\ and \U\ Stokes parameters for all the data together with the Stokes parameters for several field stars in Figure \ref{figure:quplot}.

To estimate the significance of this change, we calculated the probability that the measured values originated from one constant value, by calculating the \chisq\ values for \Q\ and \U, $\chi^2_x = \Sigma_i {\left( \frac{x_i - \overline{x}}{\sigma_{x_i}} \right)}^2$, where $x$ = \Q\ or \U, and $\overline{x}$ is the weighted mean of the 7 available measurements for the corresponding Stokes parameter. Applying an F-test to the resultant \chisq\ gives the requested probability. We also applied this procedure with $\overline{x}$ being the average ISM polarization value, and $\sigma_x$ the corresponding spread therein. 
All probabilities are small. There is a 1.2\% chance that \U\ can entirely be attributed to ISM polarization, but this probability for \Q\ is almost zero, as is then the probability for $P$. The probability that either \U\ or \Q\ belongs to one average value is less than $10^{-6}$.

\section{Discussion}

Several models explain changes in the polarization angle in the context of jetted outflow of the gamma-ray burst ejecta. The break seen in the light curve of \grb\ \citep{holland:2003,bersier:2003b} is indicative of such a jetted outflow, though the many bumps in the light curve hamper the detection of such a break. \citet{holland:2003} also exclude a spectral change due to the passing of the cooling frequency $\nu_c$ through the optical as a cause of this break.

We now proceed under the assumption that the change in the polarization is caused by a jetted outflow.

The model proposed by \citet{rossi:2002} explains changes in the polarization using a structured jet: a jet with brighter (and possibly faster) core surrounded by dimmer (and slower) wings, with a standard energy reservoir. However, this model does not predict a change in the position angle and is thus ruled out by our measurements for this burst. The model suggested by \citet{sari:1999} and \citet{gl:1999} predicts a 90 degrees change in the position angle, roughly around the time of the break. The precise moment of this change, as well as the maximum observable polarization, depends on the ratio of $\theta/\theta_0$, where $\theta$ is the angle between the jet axis and the line of sight, and $\theta_0$ is the initial opening angle of the jet. From our measurements, we deduce that the break time, where $\theta_0 \sim 1/\Gamma$ ($\Gamma$ being the bulk Lorentz factor), is somewhere between 10 hours and 1 day after the burst. The dependence of the change in polarization on $t$ is via $\Gamma$. If $\Gamma(t)$ is different than assumed in \citet{sari:1999} and \citet{gl:1999}, which is not unlikely in view of the complex behaviour of the light curve and its explanations, the time of the break will be different than estimated above. Color changes have also been seen in the optical part of the energy distribution of the burst \citep{bersier:2003b, matheson:2003}. Such behaviour requires detailed models for the polarization, whereas we have here used the general model for a smooth afterglow behaviour. It is likely, however, that these color changes are of little influence to the models described above, since they were seen past one day, while the largest change in our data is before one day. 

\citet{holland:2003} find a jet break time of $t_b = 6$ days after the burst, from fitting a broken power law to the data
. These estimates are in stark contrast with our findings. However, \citet{holland:2003} find that the break is gradual and occurred over a period of 3.5 to 7 days after the burst; they also note that their estimate of the break time might be too high, possibly putting the break around 2 days. With this latter value, our data would agree more with the jet model for polarization. The various bumps in the light curve might further obscure the detection of an early time jet break.

The smooth and gradual break in the light curve would mean that the ratio $\theta/\theta_0$ is high and we are viewing the jet close to the edge \citep[see][]{sari:1999,gl:1999}. However, this should then give rise to significantly higher values in the degree of linear polarization than measured, unless we observed close to the moment where the position angle changed by 90 degrees. That could then also explain the third set of ALFOSC observations, and would show that those observations where taken very close to the jet break time.

\section{Conclusions}

Our polarimetric observations clearly show a change in the polarization of \grb, most distinct in the position angle. The latter changes by 45\degr\ between 9 and 14 hours, and the inclusion of a later data point by \citet{covino:gcn1622} indicates even a 90\degr\ change over a 3.5 day period. Within the currently proposed GRB jet models, this would mean that we are looking at a uniform jet, with a break time of $t_b \approx 1$ days after the burst. This is in contrast with the result obtained by \citet{holland:2003}, who obtained $t_b \approx 6$ days, but with a large spread in this value ($\approx 3.5$ to $7$ days), which could still be reconciled with our findings. The structured jet model as proposed by \cite{rossi:2002} is ruled out by the fact that this model does not predict a change in the polarization angle.

\begin{acknowledgements}
ER acknowledges support from NWO grant nr. 614-51-003. JPUF
gratefully acknowledges support from the Carlsberg Foundation. This
work was supported by the Danish Natural Science Research Council
(SNF).  JMCC acknowledges the receipt of a FPI doctoral fellowship from Spain's Ministerio de Ciencia y Tecnolog\'{\i}a. The data presented here have been taken using ALFOSC, which is owned by the Instituto de Astrof\'{\i}sica de Andaluc\'{\i}a (IAA) and operated at the NOT under agreement between the IAA and the NBIfAFG of the Astronomical Observatory of Copenhagen. The NOT is operated on the island of La Palma jointly by Denmark, Finland, Iceland, Norway, and Sweden, in the Spanish Observatorio del Roque de los Muchachos of the Instituto de Astrof\'{\i}sica de Canarias. 
The FORS\,1 data were obtained as part of an ESO Service Mode run for ToO programme 70.D-0523(A). The authors acknowledge benefits from collaboration within the Research Training Network "Gamma-Ray Bursts: An Enigma and a Tool", funded by the EU under contract HPRN-CT-2002-00294.

\end{acknowledgements}

\bibliographystyle{aa}
\bibliography{Fc184}

\end{document}